\documentclass[fleqn,10pt]{wlscirep}
\usepackage[utf8]{inputenc}
\usepackage[T1]{fontenc}
\usepackage{bm,tikz,fontawesome}
\usepackage{siunitx}
\newcommand\doi[1]{\href{#1}{\,\footnotesize\textcolor{black}\faExternalLink}}

\title{Heterogeneity-induced lane and band formation in self-driven particle systems}

\def\ma{(\ref{m1})}
\def\mb{(\ref{m2})}

\author[1]{Basma Khelfa}
\author[1]{Raphael Korbmacher}
\author[2]{Andreas Schadschneider}
\author[1,*]{Antoine Tordeux}
\affil[1]{School for Mechanical Engineering and Safety Engineering, University of Wuppertal, Wuppertal, Germany}
\affil[2]{Institute for Theoretical Physics, University of Cologne, Cologne, Germany}

\affil[*]{e-mail: tordeux@uni-wuppertal.de}

\begin{abstract}
The collective motion of interacting self-driven particles describes many types of coordinated dynamics and self-organisation. 
Prominent examples are alignment or lane formation which can be observed alongside other ordered structures and nonuniform patterns. 
In this article, we investigate the effects of different types of heterogeneity in a two-species self-driven particle system.
We show that heterogeneity can generically initiate segregation in the motion 
and identify two heterogeneity mechanisms. 
Longitudinal lanes parallel to the direction of motion emerge when the heterogeneity statically lies in the agent characteristics (quenched disorder). 
While transverse bands orthogonal to the motion direction arise from dynamic heterogeneity in the interactions (annealed disorder). 
In both cases, non-linear transitions occur as the heterogeneity increases, from disorder to ordered states with lane or band patterns.
These generic features are observed for a first and a second order motion model and different characteristic parameters related to particle speed and size. 
Simulation results show that the collective dynamics occur in relatively short time intervals, persist stationary, and are partly robust against random perturbations. 
\end{abstract}

\begin{document}
\flushbottom\maketitle

\section{Introduction}

The emergence of coordinated movement and self-organisation in the collective motion of large groups of interacting autonomous individuals is ubiquitous in nature\cite{Castellano2009,Ramaswamy2010,VICSEK2012,Elgeti2015}. 
Flock and swarm behaviour arises e.g.\ in bacterial and cell migration\cite{Friedl2009, Chen2017c} (see also the pioneering works on morphogenesis by A.M. Turing\cite{Turing1952}),  
animal aggregation \cite{Vicsek1995,Parrish1999,Buhl2006,Needleman2017}, 
rods and non-living systems \cite{Narayan2007,Giomi2015,Baer2020}, 
microswimmers \cite{Janssen2017,Jahanshahi2019,Grauer2020}, 
but also in robotic systems \cite{Ibele2009,Ozkan2021}. 
This includes a large class of non-equilibrium systems of self-driven particles often called \emph{active matter} in the literature. 
The microscopic interaction rules initiate \emph{motility-induced phase separation} from disordered states to non-uniform macroscopic patterns displaying order (at least partially)\cite{Cates2015,Caprini2020}.
Pedestrian and vehicle flows are also systems of interacting self-driven agents that can exhibit phase transitions and self-organisation\cite{Chowdhury2000,Bellomo2012,Moussa2012,Shahhoseini2017,Boltes2018}. 
Indeed, pedestrians and drivers interact locally with their environment and the neighbourhood and have by the way density-dependent motility parameters. 
Similar forms of collective dynamics can arise in more abstract conceptions of self-driven particles like social systems, social networks, and opinion formation\cite{Hermann2012,Muchnik2013,Moussaid2013,Touboul2019}. 
Regardless of the important differences in the specific type and interaction of the individuals, collective motions exhibit similarities suggesting the existence of generic underlying mechanisms and systemic phase transitions\cite{Cates2015,Fu2018b,Cristin2019}. 

The terminology active, self-driven, or self-propelled particles dates back to the 1970s when it was introduced for the description of swimming micro-organisms\cite{Childress1975}. 
It became popular during the 1990s with the Vicsek model\cite{Vicsek1995} and is nowadays used for animal aggregation \cite{Buhl2006}, bacterial migration\cite{Chen2017c}, suspensions of microswimmers\cite{Jahanshahi2019}, or vehicle and pedestrian dynamics\cite{Helbing2001}. 
Self-driven particle models are mainly designed to describe collective dynamics. 
In contrast to Brownian colloids dominated by physical (exclusion) rules, the motion of self-driven particles is mainly governed by local interaction rules with the environment. 
The interactions may result from external chemical potentials, for instance in cell behaviour (chemotaxis), while they come from social and proxemics interaction rules for pedestrians (sociotaxis)\cite{Kirchner2002b}. 
The social rules of pedestrians can initiate collective phenomena improving the system performance. 
The collective motions have positive effects on transport properties and may be referred to as \emph{intelligent collective dynamics}.
Examples of intelligent collective dynamics of pedestrians are 
lane formation\cite{Burstedde2001a,Nakayama2005,Feliciani2018,Cristin2019},  
collective motion and alignment\cite{Garcimartin2017,Caprini2020}, 
or intermittent flows at bottlenecks and diagonal, chevron or circular patterns at intersections\cite{Helbing2005,Cividini2013}. 
Intelligent collective dynamics in transportation are also of practical relevance, e.g. for the control of autonomous driving systems\cite{Schwarting2019,Cerotti2017}. 
Yet, self-organisations may also induce negative effects on traffic safety and performance, for instance through stop-and-go waves\cite{Stern2018,Bain2019,friesen2021spontaneous},  
herding\cite{Kirchner2002b} or clogging effects\cite{Zuriguel2014,Nicolas2017} among other segregation phenomena\cite{Foulaadvand2007}. 
Besides scientific interests, extracting the essential features of collective motions from individual behaviors is fundamental to authorities for the control of crowd and traffic dynamics and the development of intelligent transportation strategies. 

In self-driven particle systems, collective dynamics can result from heterogeneity effects in the microscopic behaviour of the particles, upon other inertia or delay mechanisms. 
Pedestrian dynamics describe for instance lane formation for counter-flow 
or for pedestrians walking in the same direction but with different speeds\cite{Cristin2019,Fujita2019}. 
Other examples are stripe, diagonal travelling band or chevron patterns for crossing flows\cite{Helbing2005,Cividini2013}. 
In this article, we show by simulation that heterogeneity effects can generically initiate segregation and spontaneous formation of lane or band patterns in two-species flows of polarised agents. 
Two heterogeneity mechanisms are identified: static heterogeneity in the agent characteristics and dynamic heterogeneity in the interactions. 
Static heterogeneity refers to \emph{quenched disorder} in solid state physics and the terminology of random walks, when dynamic heterogeneity relies on \emph{annealed disorder} (see\cite{Krusemann2014,Tateishi2020} and references therein).
Interestingly, lanes spontaneously occur when the heterogeneity relies statically on the agent features (quenched disorder), while bands emerge if the heterogeneity operates dynamically in the interactions (annealed disorder). 
The lane and band patterns are stable and persist stationary, although no alignment interaction rules are defined (explicitly or implicitly).
The features are generically observed with different microscopic motion models, namely the first order collision-free speed model\cite{Tordeux2016} and the inertial second order social force model\cite{Helbing1995}, and different types of parameters related to agent speed or agent size. 
Lane and band patterns are observed with different binary mixtures of interacting particles \cite{lowen2010particle,poncet2017universal,vasilyev2017cooperative}, e.g.\ oppositely charged colloids subject to, respectively, DC and AC external electric fields \cite{vissers2011lane,vissers2011band}. 
In the presented models, the heterogeneity comes from internal interaction mechanisms. 
Potential applications are mixed urban traffic flow and the modelling of the interactions between different types of road users.

\subsection{Models}

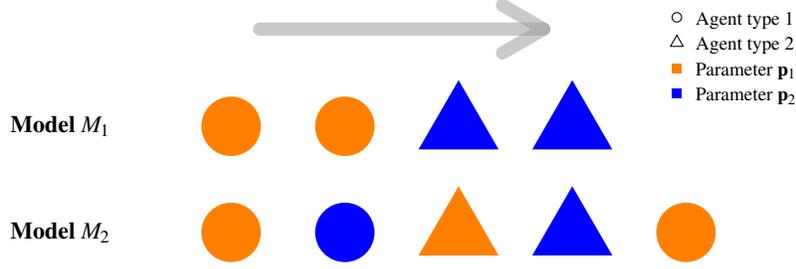
\begin{figure}[!ht]
\centering
\medskip
\begin{tikzpicture}[x=1pt,y=1pt]
\definecolor{fillColor}{RGB}{255,255,255}
\path[use as bounding box,fill=fillColor,fill opacity=0.00] (0,0) rectangle (325.21,108.41);
\begin{scope}
\path[clip] (  0.00,  0.00) rectangle (325.21,108.41);
\definecolor{fillColor}{RGB}{255,128,0}

\path[fill=fillColor] ( 98.08, 57.55) circle ( 11.25);

\path[fill=fillColor] (141.10, 57.55) circle ( 11.25);
\definecolor{fillColor}{RGB}{0,0,255}

\path[fill=fillColor] (184.12, 75.04) --
	(199.27, 48.80) --
	(168.97, 48.80) --
	cycle;

\path[fill=fillColor] (227.13, 75.04) --
	(242.29, 48.80) --
	(211.98, 48.80) --
	cycle;
\definecolor{fillColor}{RGB}{255,128,0}

\path[fill=fillColor] ( 98.08, 17.40) circle ( 11.25);
\definecolor{fillColor}{RGB}{0,0,255}

\path[fill=fillColor] (141.10, 17.40) circle ( 11.25);
\definecolor{fillColor}{RGB}{255,128,0}

\path[fill=fillColor] (184.12, 34.89) --
	(199.27,  8.65) --
	(168.97,  8.65) --
	cycle;

\definecolor{fillColor}{RGB}{0,0,255}
\path[fill=fillColor] (227.13, 34.89) --
	(242.29,  8.65) --
	(211.98,  8.65) --
	cycle;
\definecolor{fillColor}{RGB}{220,220,220}
\definecolor{fillColor}{RGB}{255,128,0}

\path[fill=fillColor] (270.15, 17.40) circle ( 11.25);

	cycle;
\definecolor{drawColor}{RGB}{0,0,0}

\node[text=drawColor,anchor=base,inner sep=0pt, outer sep=0pt, scale=  0.90] at ( 33.55, 55.22) {\textbf{Model $M_1$}};

\node[text=drawColor,anchor=base,inner sep=0pt, outer sep=0pt, scale=  0.90] at ( 33.55, 15.07) {\textbf{Model $M_2$}};

\path[draw=drawColor,line width= 0.4pt,line join=round,line cap=round] (266.68, 98.81) circle (  1.80);

\path[draw=drawColor,line width= 0.4pt,line join=round,line cap=round] (266.68, 92.00) --
	(269.10, 87.81) --
	(264.26, 87.81) --
	(266.68, 92.00);
\definecolor{fillColor}{RGB}{255,128,0}

\path[fill=fillColor] (264.88, 77.81) --
	(268.48, 77.81) --
	(268.48, 81.41) --
	(264.88, 81.41) --
	cycle;
\definecolor{fillColor}{RGB}{0,0,255}

\path[fill=fillColor] (264.88, 68.21) --
	(268.48, 68.21) --
	(268.48, 71.81) --
	(264.88, 71.81) --
	cycle;

\node[text=drawColor,anchor=base west,inner sep=0pt, outer sep=0pt, scale=  0.80] at (273.88, 96.05) {\small Agent type $1$};

\node[text=drawColor,anchor=base west,inner sep=0pt, outer sep=0pt, scale=  0.80] at (273.88, 86.45) {\small Agent type $2$};

\node[text=drawColor,anchor=base west,inner sep=0pt, outer sep=0pt, scale=  0.80] at (273.88, 76.85) {\small Parameter $\mathbf p_1$};

\node[text=drawColor,anchor=base west,inner sep=0pt, outer sep=0pt, scale=  0.80] at (273.88, 67.25) {\small Parameter $\mathbf p_2$};
\definecolor{drawColor}{RGB}{150,150,150}

\path[draw=drawColor,draw opacity=0.50,line width= 5.0pt,line join=round,line cap=round] (108.84, 94.35) -- (216.38, 94.35);

\path[draw=drawColor,draw opacity=0.50,line width= 5.0pt,line join=round,line cap=round] (200.73, 85.32) --
	(216.38, 94.35) --
	(200.73,103.39);
\end{scope}
\end{tikzpicture}
\medskip
\caption{Illustrative scheme in one dimension for the two heterogeneity models. The parameter setting (orange or blue) depends on the type of agents (represented as disc and triangle) for the model Eq.~\protect\ma\ while the setting depends on the type of the agent in front for the model Eq.~\protect\mb: it is orange when the agent in front is of the same type and blue if it is of another type.}
\label{fig1}
\end{figure}

We consider in the following two types of agents evolving on a torus. 
We denote $n=1,\ldots,N$ the agent's ID while $k_n={1,2}$ is the agent's type.
The agent's motion is given by a dynamic model $F_{\mathbf p}(\mathbf X_n)$ 
that defines the agent speed as in the collision-free model\cite{Tordeux2016} or the agent acceleration as in the social force model\cite{Helbing1995} according to local spatio-temporal variables $\mathbf X_n$ (e.g.\ the position and speed differences with the neighbours) and a set of parameters $\mathbf p$ (namely, desired speed, desired time gap, repulsion rate, agent size, and so on). 
We assume two different settings $\mathbf p_1$ and $\mathbf p_2$ for the parameters. 
Two types of heterogeneity are then considered.
\begin{enumerate}
    \item {\emph{Heterogeneity in the agent characteristics}} -- 
		We attribute statically the two parameter settings $\mathbf p_1$ and $\mathbf p_2$ to the two types of agents:
		\begin{equation}
		M_1(n,k_n)=F_{\mathbf p_{k_n}}(\mathbf X_n).
        \label{m1}
		\end{equation}
		We aim here to model different types of agents (for instance pedestrians and bicycles) with specific characteristics in term of desired speed, agent size, etc. 
		This kind of heterogeneity is usually called \emph{quenched disorder} in solid state physics. 
		It refers to static heterogeneity features remaining constant (i.e.\ quenched) over the time.
    \item {\emph{Heterogeneity in the interactions}} -- 
		We attribute dynamically the two parameter settings  $\mathbf p_1$ and $\mathbf p_2$ according to the type of the closest agent in front. The parameter setting is $\mathbf p_1$ if the agent in front is of the same type, 
		while it is $\mathbf p_2$ in case of interaction with an another agent type:
				\begin{equation}
		M_2(n,k_n)=\left\{\begin{array}{ll}F_{\mathbf p_1}(\mathbf X_n),&\text{if $\tilde k(\mathbf X_n)=k_n$,}\\[1mm]
		F_{\mathbf p_2}(\mathbf X_n),&\text{otherwise,}\end{array}\right.
        \label{m2}
		\end{equation}
		with $\tilde k(\mathbf X_n)$ the type of the closest agent in front (see Sec.~\ref{meth} for details). 
		Such a mechanism may be realized in mixed urban traffic where cyclists or electric scooter drivers are adapting their behaviour, increasing the time gap or reducing his/her desired speed, when following a group of pedestrians. 
		The heterogeneity features are here time-dependent. 
		They are usually called \emph{annealed disorder} in the literature of solid state physics\cite{Krusemann2014,Tateishi2020}.
\end{enumerate}
In contrast to the model Eq.~\ma\ for which the heterogeneity statically lies in agent characteristics, the model Eq.~\mb\ induces a dynamic heterogeneity mechanism taking place in the interactions. 
See Fig.~\ref{fig1} for an illustrative example in one dimension.

\subsection{Analysis}

We qualitatively observe by simulation that the static heterogeneity model $M_1$ Eq.~\ma\ initiates the formation of lanes in the system, while the dynamic heterogeneity model $M_2$ Eq.~\mb\ allows the formation of bands (see Fig.~\ref{fig2} below).
To classify the state of the system, we measure the agent's mean speed and also order parameters for the lane and band formation. 
The order parameter has been introduced to detect lanes in a colloidal suspension\cite{Rex2007} and used in pedestrian dynamics\cite{Nowak2012}. 
We denote in the following $(x_n,y_n)$ the positions of the agents $n=1,\ldots,N$. 
The order parameter for lane formation is
\begin{equation}
\Phi_L=\frac1N\sum_n\phi_n^L\qquad\text{with}\qquad
\phi^L_n=\Big[\frac{L_n-\bar L_n}{L_n+\bar L_n}\Big]^2
\qquad\left|\begin{array}{l}
L_n=\text{card}\big(m,\,|y_n-y_m|<\Delta,~k_n=k_m\big)\\[1mm]
\bar L_n=\text{card}\big(m,\,|y_n-y_m|<\Delta,~k_n\ne k_m\big)\end{array}\right..
\end{equation}
Here $L_n$ is the number of agents on a lane of width $\Delta>0$ in front of the agent $n$ with the same type, $\text{card}(A)$ being the operator counting the elements of an ensemble $A$, while $\bar L_n$ is the number of agents with different types. 
The order parameter $\Phi_L$ tends by construction to be close to one when the system describes lanes.
Assuming a disordered state for which the agents are uniformly randomly distributed on a $w\times h$ rectangle with $h>\Delta>0$ the system's height and $w>0$ the system's width, the number $L_n$ of agents with the same type is distributed according to the binomial model  $\mathcal B(m,p)$, with $m=N_{k_n}$ and $p=\Delta/h$. Here $N_{k_n}$ is the total number of agents with type $k_n$.
The distribution of the number $\bar L_n$ of agents with different types can be deduced similarly. 

For band formation, the order parameter is
\begin{equation}
\Phi_B=\frac1N\sum_n\phi_n^B\qquad\text{with}\qquad
\phi^B_n=\Big[\frac{B_n-\bar B_n}{B_n+\bar B_n}\Big]^2\qquad
\left|\begin{array}{l}B_n=\text{card}\big(m,\,|x_n-x_m|<\Delta w/h,~k_n=k_m\big)\\[1mm]
\bar B_n=\text{card}\big(m,\,|x_n-x_m|<\Delta w/h,~k_n\not=k_m\big)\end{array}\right..
\end{equation}
The band order parameter includes a term $w/h$, $w$ and $h$ being the width and height of the system. 
The distribution of the order parameters for lanes and bands is by construction the same in cases of random positions of the agents.
Indeed for disordered states, the number $B_n$ of agents on the sides with the same type has a binomial distribution $\mathcal B(m,p)$ with $m=N_{k_n}$ and $p=\Delta w/(hw)=\Delta/h$ as well. 
This makes the lane and band order parameters directly comparable. 
In particular, $\mathbb E\Phi_L=\mathbb E\Phi_B\ge (1-p)/(1-p+pm)$ systematically holds for disordered states.  

\section{Simulation results}

We carry out simulations of two-species flows on a $9\times 5$\,m rectangular with top-down and right-left periodic boundary conditions (torus). 
We simulate the evolution of $N=45$ agents (density of $1$~agent$/$m$^2$) from random initial conditions using the first order collision-free (CF) pedestrian model\cite{Tordeux2016} and the inertial social force (SF) model\cite{Helbing1995} in the Supplementary Materials. 
The desired directions of motion of all agents are polarised to the right.  
The heterogeneity in the two settings $\mathbf p_1$ and $\mathbf p_2$ is introduced by varying model parameters related to the speed (i.e.\ desired speed or time gap parameters) or to the size of the agents. 
We quantify the heterogeneity level in the two-species system using the index $\delta_s$ when we vary parameters related to agent speed, and the index $\delta_l$ when we vary parameters related to agent size.
The definitions of the microscopic motion model and details on the setting of the model's parameters and heterogeneity indexes are provided in Sec.~\ref{meth}. 

\subsection{Preliminary experiment}

We first present single simulation histories of the two-species system with the two heterogeneity models Eq.~\ma\ and Eq.~\mb. 
We simulate the evolution of the agents using the collision-free model\cite{Tordeux2016} for given heterogeneity indexes $\delta_s$ on the parameters related to the agent speed, namely the desired speed and the time gap parameters (see Sec.~\ref{meth} for details on the setting of the model parameters). 
Successive snapshots of the system are presented in Fig.~\ref{fig2}. 
The evolution of the system with the static heterogeneity model Eq.~\ma\ is shown in the left panels while the evolution with the dynamic heterogeneity model Eq.~\mb\ is displayed in the right panels. 
The bottom panels provide the evolution of the order parameters for lane and band formation. 
We observe fast formation of two lanes by agent type within the first heterogeneity model, while two bands emerge with the second model. 
The parameter settings are statically attributed to the agent type for the model defined by Eq.~\ma. Thus, the segregation also involves the parameter setting. 
In contrast, the parameter setting depends on the type of the agent in front for the model defined by Eq.~\mb. 
This results in four bands according to the parameter settings.
Note that further simulations with larger systems may describe more lanes and bands with different sizes. 

\newcommand\addpl[5]{
{\footnotesize$t=#1,\quad~\bar v=#2\text{\,m/s},\quad~ \Phi_L=#3,\quad~ \Phi_B=#4$}\\[1mm]
\includegraphics[width=\textwidth]{#5}\\[3mm]}
\begin{figure}[!ht]\centering\bigskip\medskip
\centering 
\begin{minipage}[t]{.38\textwidth}
\begin{center}
\textbf{\small Model $M_1$} -- Static heterogeneity
\end{center}
\addpl{0}{0.63}{0.23}{0.14}{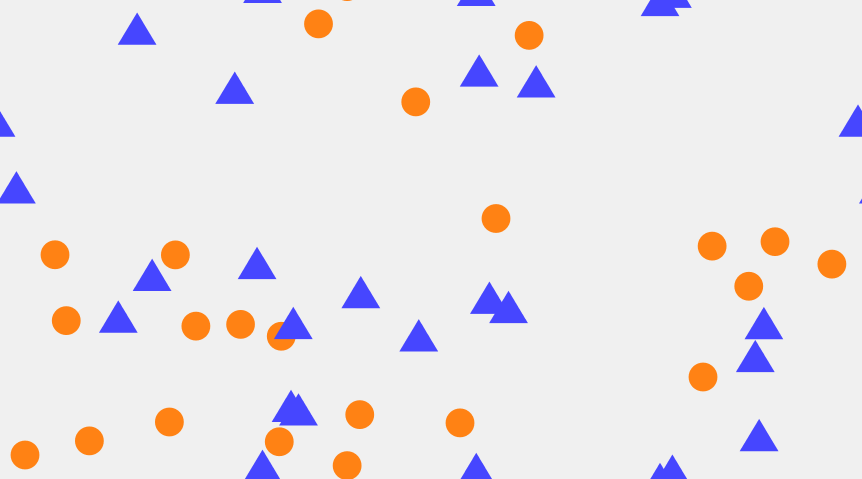}
\addpl{10}{1.36}{0.51}{0.24}{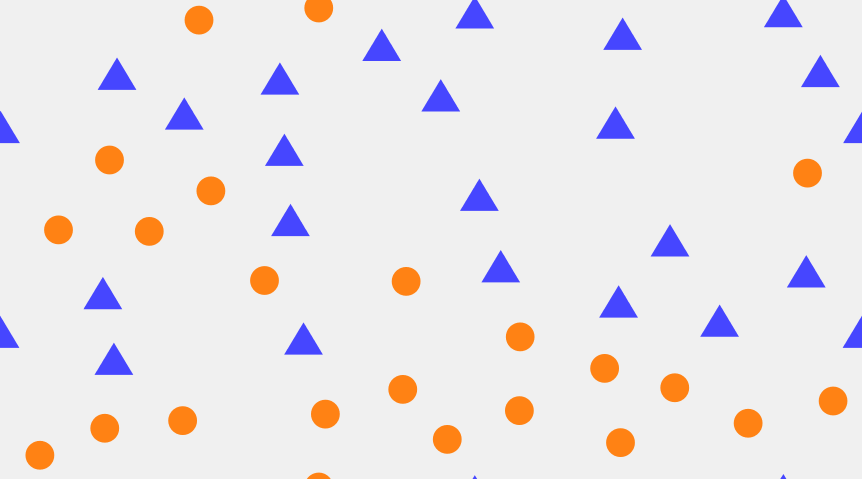}
\addpl{50}{1.35}{0.97}{0.23}{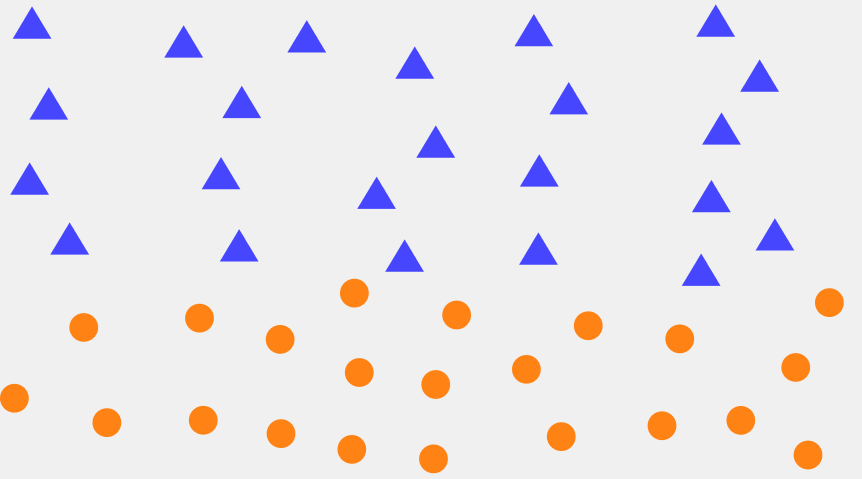}
\end{minipage}\hspace{2cm}\begin{minipage}[t]{.38\textwidth}
\begin{center}
\textbf{\small  Model $M_2$} -- Dynamic heterogeneity
\end{center}
\addpl{0}{0.51}{0.19}{0.29}{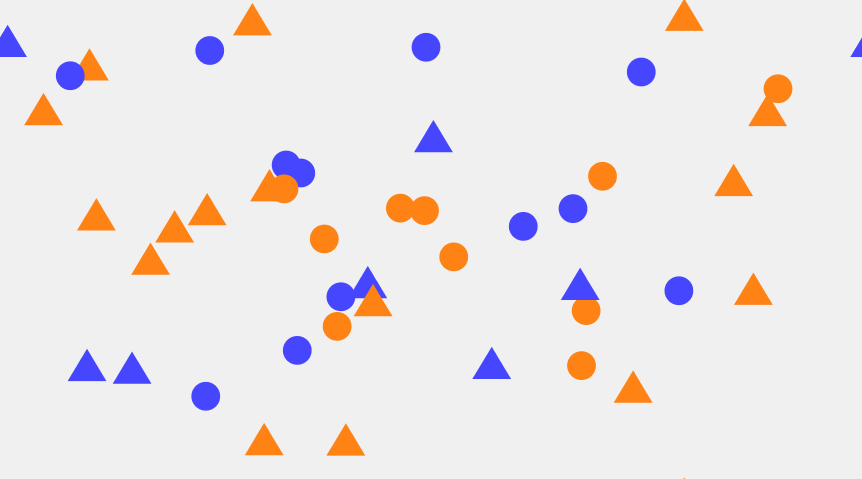}
\addpl{10}{1.22}{0.13}{0.5}{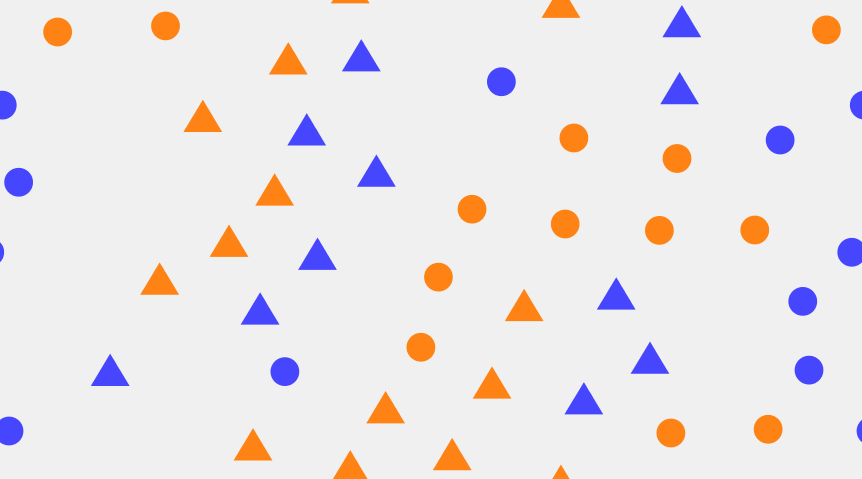}
\addpl{50}{1.28}{0.12}{1}{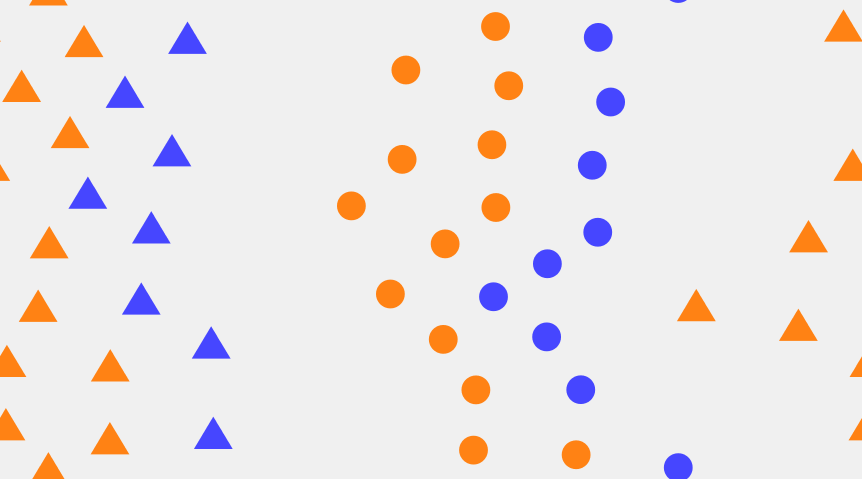}
\end{minipage}\\[-2mm]
\input{Figures/fig2g.tex}
    \caption{Typical histories for the model Eq.~\protect\ma\ with heterogeneity in the agent characteristics for which lanes emerge: the lane order parameter tends to one while the band order parameter is close to zero (left panels, $\delta_s=18$), and for the model Eq.~\protect\mb\ with heterogeneity in the interactions where bands emerge with opposite characteristics for the order parameter (right panels, $\delta_s=8$). Flow direction from left to right, periodic boundary conditions, random initial conditions.}
    \label{fig2}
\end{figure}

The order parameters converge after a transient phase to stationary performances with lanes or bands where they are polarised to one or zero. 
The duration of the transient states is approximately 40 seconds of simulation. 
Note that the duration of the transient states varies from a simulation to another but the system systematically converges to a stationary state with lanes or bands. 
Furthermore, lane and band formation in larger systems require longer simulation times, especially for the band formation (see the blue dotted curves in Fig.~\ref{fig2}, bottom panel, for a $15\times9$\,m system three times larger with $135$ pedestrians).
Similar performances are observed when using the social force model instead of the collision-free model (see Fig.~S1 in the Supplementary Materials).
Here, the heterogeneity of the two parameter settings $\mathbf p_1$ and $\mathbf p_2$ and corresponding index $\delta_s$ are relatively high. 
Reducing the heterogeneity index can result in a longer transient phase or even no formation of lanes and bands. 
We may expect that lanes and bands progressively emerge as the heterogeneity index increases. 
This is however not the case.
As described in the next section, we observe in stationary states an abrupt phase transition from disorder states to order states with lanes or bands as the heterogeneity index increases.

\subsection{Stationary performances}

The preliminary experiment shows that lanes tend to emerge in the dynamics when the heterogeneity relies on agent characteristics (quenched disorder model $M_1$ Eq.~\ma) while bands occur when the heterogeneity takes place in the interactions (annealed disorder model $M_2$ Eq.~\mb). 
The results presented in Fig.~\ref{fig2} are obtained for given values of the heterogeneity index $\delta_s$ between the two parameter settings $\mathbf p_1$ and $\mathbf p_2$. 
The index is sufficiently high to rapidly observe the formation of lanes or bands.
In this section, we analyse the performances by progressively increasing the heterogeneity indexes $\delta_s$ and $\delta_l$. 
We repeated one thousand Monte-Carlo simulations from independent random initial configurations for the two heterogeneity models $M_1$ Eq.~\ma\ and $M_2$ Eq.~\mb\ by varying the heterogeneity indexes $\delta_s$ and $\delta_l$ over twenty levels. 
The differences between the two parameter settings $\mathbf p_1$ and $\mathbf p_2$ are zero at the lower heterogeneity level, while they are important at the higher level. 
We measure the system during $60$\,s after a simulation time $t_0=600$\,s to observe stationary performances.
The remaining Figs.~\ref{fig-cfmspeed}--\ref{fig-cfmsize} and Figs.~S2--S5 in the Supplementary Materials show the median estimates of the Monte-Carlo simulations with inter-quartile ranges of the agent mean speed and order parameters for lane and band formation.
Details on the setting of the model's parameters and heterogeneity indexes are provided in Sec.~\ref{meth}. 

We first present the Monte-Carlo simulation results obtained by varying the heterogeneity index $\delta_s$ relying on agent speed features (desired speed and time gap parameters).
We observe the emergence of lanes when the heterogeneity relies on agent characteristics (quenched disorder model $M_1$ Eq.~\ma, see Fig.~\ref{fig-cfmspeed}, left panels) while band occurs when the heterogeneity operates in the interactions (annealed disorder model $M_2$ Eq.~\mb, see Fig.~\ref{fig-cfmspeed}, right panels). 
An abrupt phase transition occurs as the heterogeneity index $\delta_s$  increases from a disordered state for which the order parameters are close to 0.2 (dotted line in Fig.~\ref{fig-cfmspeed}, top panels) to an ordered dynamics with lanes or bands for which the order parameters are polarised on zero or one. 
A critical heterogeneity index can be identified.
The lane patterns allow the speed of the agents with faster characteristics to be higher than the speed of agents with slower features (Fig.~\ref{fig-cfmspeed}, bottom left panel). 
This makes the agent speed on average close to the mean speed of a homogeneous flow (dotted line). 
In contrast, the band patterns in the model $M_2$ Eq.~\mb\ correspond to gridlocks for which the speed of all the agents have slower features (Figs.~\ref{fig-cfmspeed}, bottom right panel).
Similar performances occur by varying parameters relying on agent size (see Fig.~\ref{fig-cfmsize} below), or by using the social force inertial model instead of the first order collision-free model (see Figs.~S2 and S3 in the Supplementary Materials).

The lane formation for counter-flow is an universal collective mechanism observed in pedestrian dynamics and binary mixtures of interacting particles \cite{Burstedde2001a,Nakayama2005,vissers2011lane,Feliciani2018,poncet2017universal,vasilyev2017cooperative,Cristin2019}. 
It relies on heterogeneity of agent velocity: some of them have a desired speed $v$ while the others have a desired speed $-v$.
No counter-flow occurs in the presented simulation experiments since all agents are polarised to the right. 
The agents of type 1 have a speed $v_1>0$ and agents of type 2 have a speed $v_2>0$ with $v_1\not=v_2$. 
However, using a Galilean transformation with $\tilde v=(v_1+v_2)/2$ and neglecting anisotropic effects the situation is the same as in the counter-flow with $v=|v_2-v_1|/2$. 
Therefore, it is not surprising to observe longitudinal lane formation in uni-directional flows with the quenched disorder model $M_1$ Eq.~\ma\ even if no counter-flow arises (see also recent experiments\cite{Fujita2019}). 
The dynamic heterogeneity features of the annealed disorder model $M_2$ Eq.~\mb\ result in transversal segregation in space and the formation of bands vertical to the direction of motion. 
Band structures, as clusters or stop-and-go dynamics, are self-organised density waves \cite{vissers2011lane,Bain2019}.
The agents have a constant speed and the bands propagate downstream in the presented study.
The bands propagate upstream with a characteristic speed and enforce agents to deceleration and acceleration phases in case of stop-and-go dynamics \cite{Nakayama2008,Stern2018,Bain2019} while they are static for oppositely charged particles \cite{vissers2011band}. 

\begin{figure}[!ht]
\centering\smallskip
\input{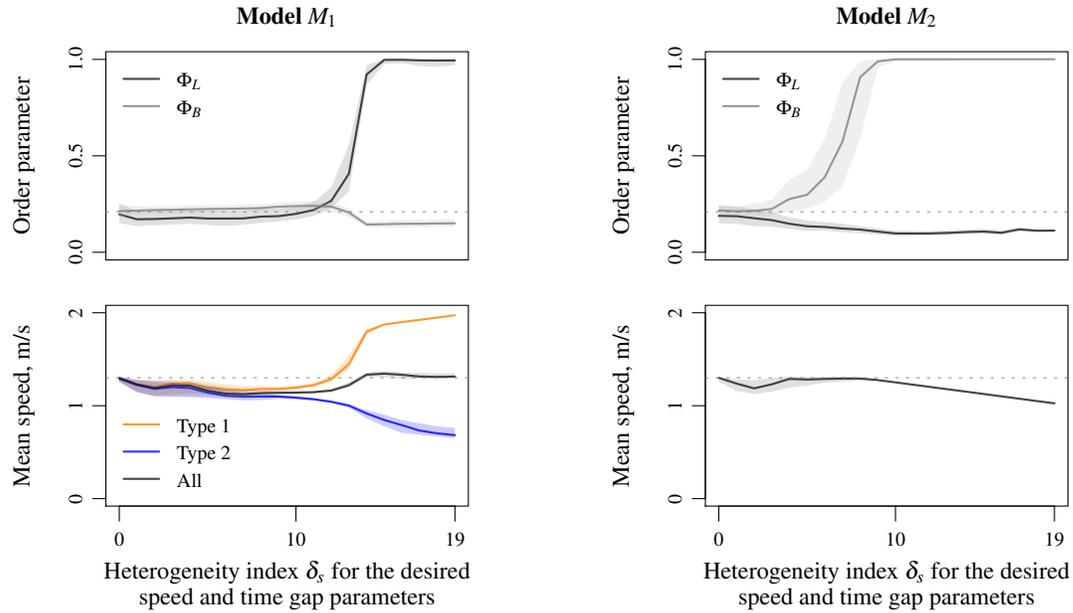}
\caption{Lane and band order parameters (top panels) and mean speed (bottom panels) according to the heterogeneity index $\delta_s$ for agent speed features. 
Longitudinal lanes emerge with the model $M_1$ Eq.~\protect\ma\ for which the heterogeneity relies on agent characteristics: $\Phi_L$ tends to $1$ while $\Phi_B$ is low (top left panel). 
Orthogonal bands appear with the model $M_2$ Eq.~\protect\mb\ for which the heterogeneity operates in the interaction: $\Phi_B$ tends to $1$ while $\Phi_L$ is low (top right panel). 
The lane pattern allows the speed of the agents with faster characteristics to be higher than the speed of agents with slower features (bottom left panel). 
In contrast, the band pattern acts as a gridlock for which all the agents have slower speed features (bottom right panel).}
\label{fig-cfmspeed}
\end{figure}

\begin{figure}[!ht]
\centering
\input{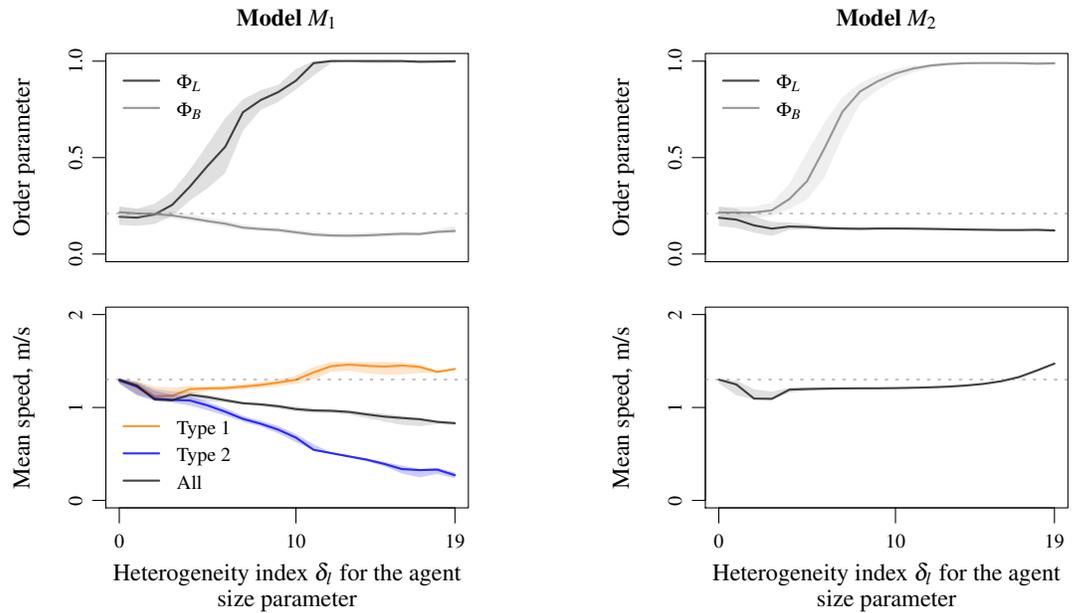}
\caption{Lane and band order parameters (top panels) and mean speed (bottom panels) according to the heterogeneity index $\delta_l$ for agent size. Comparable although smoother transitions to patterns with lanes or bands arise when varying parameters rela- ted to agent size (compare to Fig.~\protect\ref{fig-cfmspeed}) suggesting universal characteristics of the heterogeneity mechanisms Eq.~\protect\ma\ and Eq.~\protect\mb.}
\label{fig-cfmsize}
\end{figure}

Heterogeneity in agent speed featuring results in transitions from disordered states to ordered patterns with lanes or bands according to, respectively, model $M_1$ Eq.~\ma\ and model $M_2$ Eq.~\mb.
Similar although smoother transitions occur when varying the agent size characteristic index $\delta_l$ instead of the speed index $\delta_s$ (compare Figs.~\ref{fig-cfmspeed} and \ref{fig-cfmsize}). 
Lanes emerge for the static heterogeneity model Eq.~\ma, see Fig.~\ref{fig-cfmsize}, left panels, while bands arise for the dynamic heterogeneity model Eq.~\mb, see Fig.~\ref{fig-cfmsize}, right panels.
In contrast to heterogeneous models relying on agent speed, varying the agent size induces bi-dimensional steric effects making the average speed in the presence of lanes less than the mean speed of a homogeneous flow (see Fig.~\ref{fig-cfmsize}, bottom left panel). 
On contrast, the mean speed can be higher than the homogeneous one in the presence of bands (see Fig.~\ref{fig-cfmsize}, bottom right panel).
Indeed, varying the agent size acts in two dimensions, reducing or increasing the available space in case of presence of lanes or bands. 
Similar performances occur when using the social force model instead of the collision-free model (compare Fig.~\ref{fig-cfmsize} and Fig.~S3 in the Supplementary Materials).

\subsection{Transient states and perturbed systems\label{trans}}

\begin{figure}[!ht]
\centering
\input{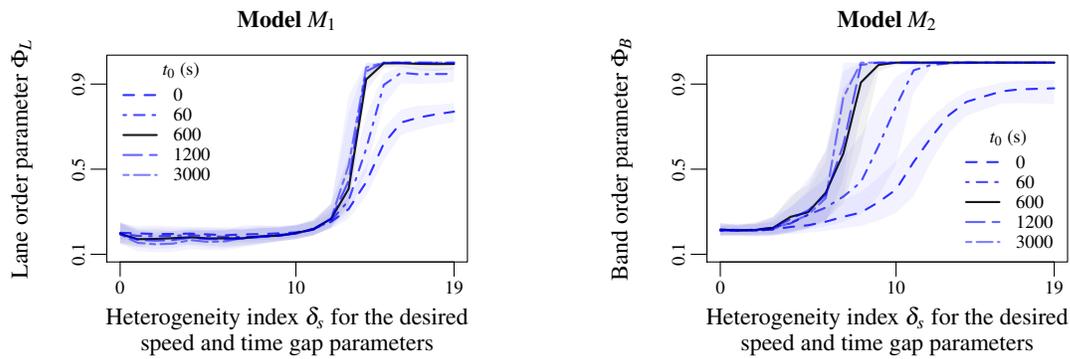}\smallskip
\caption{Lane order parameter for the model $M_1$ Eq.~\protect\ma\ (left panel) and band order parameter for the model $M_2$ Eq.~\protect\mb\ (right panel) according to the heterogeneity index $\delta_s$ of agent speed features. 
The different curves correspond to different simulation times $t_0=0$, $60$, $600$, $1200$ and $3000$\,s before starting the measurements (random initial conditions).
The phase transition to lane and band patterns emerges relatively quickly. It can be observed during the first minutes of simulation. 
More precisely, the order parameters are similar for $t_0=600$, $1200$ and $3000$\,s suggesting that the system is stationary as soon as $t\ge600$\,s.}
\label{fig-cfmspeed-difft0}
\end{figure}

The simulations above describe stationary situations.
Yet, it is interesting to observe the transient states of the system and the time required for the emergence of lanes or bands. 
In Fig.~\ref{fig-cfmspeed-difft0}, we run simulations for different simulation times $t_0=0$, $60$, $600$, $1200$ and $3000$\,s before starting the measurements. 
The initial conditions are random. 
The lanes and bands spontaneously emerge during the first minute of simulation when the heterogeneity index $\delta_s$ is sufficiently high. 
Similar phase transition to lane and band patterns occur for $t_0=600$, $t_0=1200$ and $t_0=3000$\,s suggesting that the dynamics can be considered stationary as soon as $t\ge600$\,s. 
The simulation times required to obtain stationary performances fluctuate from one simulation to another. 
They also depend on the size of the system and the density level. 
Generally speaking, larger or more dense systems require on average longer simulation times to reach a stationary state than smaller or least dense systems.

\begin{figure}[!ht]
\centering
\input{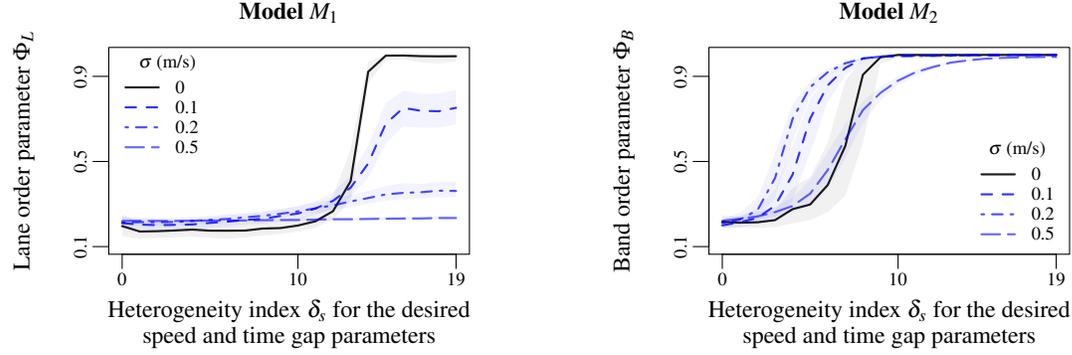}
\caption{Lane order parameter for the model $M_1$ Eq.~\protect\ma\ (left panel) and band order parameter for the model $M_2$ Eq.~\protect\mb\ (right panel) according to the heterogeneity index $\delta_s$ of agent speed features. 
The different curves correspond to different noise amplitudes $\sigma=0$, $0.1$, $0.2$ and $0.5$\,m/s in the dynamics.
The noise clearly perturbs the lane formation (left panel). Such a phenomenon is well-known in the literature as freezing-by-heating-effect\protect\cite{Helbing2000a}.
Oppositely, the noise initially improves the band formation for $\sigma=0.1$ or $0.2$\,m/s (noise-induced-ordering-effect), before partly altering it for $\sigma=0.5$\,m/s (right panel).}
\label{fig-cfmspeed-diffsig}
\end{figure}

So far, the modelling approach is deterministic. 
Analysing whether the collective motion is robust against random noising may reveal unexpected behaviours. 
In Fig.~\ref{fig-cfmspeed-diffsig}, we present the order parameter for stochastic systems for which the agent speeds are subject to independent Brownian noises. 
Simulations are carried out for a noise amplitude $\sigma=0.1$, $0.2$ and $0.5$\,m/s. 
The noise monotonically perturbs the lane formation in the static heterogeneity model $M_1$ Eq.~\ma\ (Fig.~\ref{fig-cfmspeed-diffsig}, left panel). 
No phase transition occurs for $\sigma=0.5$\,m/s. 
This phenomenon is well known in the literature as the freezing-by-heating effect \cite{Helbing2000a}. 
The concept is borrowed from the plant growth stimulation process. 
On the contrary, introducing a low noise in the dynamics allows improving the band formation in the dynamic heterogeneity model $M_2$ Eq.~\mb. 
Indeed, the critical heterogeneity indexes for the phase transition are smaller with $\sigma=0.1$ or $\sigma=0.2$\,m/s than for the deterministic model with $\sigma=0$, see Fig.~\ref{fig-cfmspeed-diffsig}, right panel. 
The band pattern is not only robust to the noise, it exists noise ranges improving the band formation. 
In contrast to the freezing-by-heating effect, this can be interpreted as an example of  \emph{noise-induced ordering effect} \cite{wiesenfeld1994,dhuys2021}.

\section{Discussion}

Two microscopic heterogeneity mechanisms are identified for the formation of collective segregation in two-species systems of polarised agents.
Quenched disorder model $M_1$ Eq.~\ma\ with static heterogeneity in the agent characteristics initiates the formation of longitudinal lanes. 
While annealed disorder model $M_2$ Eq.~\mb\ for which the heterogeneity dynamically lies in the interactions initiates transversal bands perpendicular to the motion direction. 
These features are observed when varying parameters related to agent speed (e.g.\ desired speed or time gap parameters, see Figs.~\ref{fig2} and \ref{fig-cfmspeed}) or to agent size (see Fig.~\ref{fig-cfmspeed}), and by using the first order collision-free model\cite{Tordeux2016} or the second order social force model\cite{Helbing2000a} (see Fig.~S1--S3 in the Supplementary Materials). 
The lanes and bands already emerge rather early in the simulations (see Figs.~\ref{fig-cfmspeed-difft0} and S4 in the Supplementary Materials) and are partly robust against random perturbations (see Figs.~\ref{fig-cfmspeed-diffsig} and S5 in the Supplementary Materials). 
The band formation is especially robust and may even be improved by perturbing the agent dynamics (noise-induced ordering effect), when the lane formation does not (freezing-by-heating effect). 
Further simulation results show similar behaviours when the noise is introduced in the agent polarity (i.e.\ the desired direction) instead of the speed. 

Regardless of the modelling order of the motion models (speed-based or acceleration-based models) and related parameters, the static quenched disorder and dynamic annealed disorder heterogeneity mechanisms result in collective formation of lanes or bands. 
Relying the heterogeneity on agent characteristics or on the interactions initiates generically segregation and the formation of lanes and bands in the dynamics.
Such results corroborate the universality of the lane formation observed in pedestrian counter-flows, oppositely charged colloids, upon other binary mixtures of interacting particles \cite{vissers2011lane,poncet2017universal,vasilyev2017cooperative,Cristin2019}. They open new explanation perspectives for the formation of bands. 
Further theoretical investigations remain necessary to rigorously characterise the phase transitions. 
A possibility is to analyse mean-field instability phenomena of discrete lattice representations of the model \cite{Cividini2013}. 
The presence of walls and obstacles and the role of the geometry of given facilities may also be of interest. 
Preliminary simulation results show segregation effects of slower or bigger agents in case of bottleneck within the static heterogeneity model. 
These are expelled at the edges of the system and are obstructed by the presence of walls.
These simulation results require more attention, notably for elderly people, people with motor disabilities, or in the current context of social distanciation.

\section{Methods\label{meth}}

The two agent motion models used in the simulations are the collision-free (CF) model\cite{Tordeux2016} and the social force (SF) model\cite{Helbing1995} (see the Supplementary Materials).
The CF model is a speed-based model of first order while the SF model is an inertial acceleration-based model. 
The agent dynamics are polarised in both models, i.e.\ all the agents have identical desired direction.

\paragraph{Collision-free model}

In the collision-free model, the dynamics of an agent $n$ with position $\mathbf x_n$ is given by the first order differential equation
\begin{equation}
\dot{\mathbf x}_n= F^\textsc{\tiny cf}_{\mathbf p_j}(\mathbf X_n)=V(\mathbf X_n,p_j)\;\mathbf e(\mathbf X_n,p_j)+\sigma\xi_n,
\label{cfm}
\end{equation}
composed of the scalar speed model
\begin{equation}
V(\mathbf X_n,p_j)=\max\big\{0,\min\big\{V_j,\big(s(\mathbf X_n)-\ell_j\big)/T_j\big\}\big\},
\end{equation}
here $V_j\ge0$ is the desired speed, $T_j>0$ denotes the desired time gap, and $\ell_j\ge0$ the agent size, the index $j=1,2$ representing the two parameter setting $\mathbf p_1$ and $\mathbf p_2$, and the direction model
\begin{equation}
\mathbf e(\mathbf X_n,p_j)=\frac1C\Big(\mathbf e_0+\sum_{m\not=n}\nabla U(\|\mathbf x_n-\mathbf x_m\|)\Big),
\end{equation}
with $\mathbf e_0=0$ the desired direction (polarity), $U(x)=A\exp\big((\ell_j-x)/B\big)$ with parameters $A=5$ and $B=0.1$\,m a repulsive potential with the neighbors, and $C>0$ a normalisation constant. A bi-dimensional white noise $\xi_n$ (i.e. the time derivatives of two independent Wiener processes) with amplitude $\sigma>0$ is used for the stochastic model in Sec.~\ref{trans}. 
The function
$s(\mathbf X_n)=\|\mathbf x_n-\mathbf x_{m_0(\mathbf X_n)}\|$ in the scalar speed model determines the minimal distance in front,
\begin{equation}
m_0(\mathbf X_n)=\text{arg}\min_{m\in N_n} \|\mathbf x_n-\mathbf x_m\|,\qquad N_n=\big\{m,~\mathbf e_n\cdot\mathbf e_{nm}\le0\ \mbox{ and }\ |\mathbf e_n^\perp\cdot\mathbf e_{nm}|\le\ell/\|\mathbf x_n-\mathbf x_m\|\big\},
\label{m0}
\end{equation}
being the closest agent in front of the agent $n$.
Note that in the definition of the dynamic heterogeneity type Eq.~\mb, the type of the closest agent in front is $\tilde k(\mathbf X_n)=k_{m_0(\mathbf X_n)}$.
The simulations are carried out using an explicit Euler numerical scheme in deterministic cases, and using an Euler-Maruyama scheme for the simulations including the stochastic noise. 
The time step is $\delta t=0.01$\,s in both cases. 

\paragraph{Setting of the parameters}

The default values for the parameters $\mathbf p=(\ell,V,T)$ of the CF model are based on the setting proposed in the literature\cite{Tordeux2016}
\begin{equation}
\ell=0.3~\text{m},\quad V=1.5~\text{m/s},\quad T=1~\text{s}.
\end{equation} 
Note that $\Delta=0.6$\,m in the order parameters corresponds approximately to two times the size of a pedestrian.
Starting from the default values, we vary using heterogeneity indexes the parameter settings $\mathbf p_1=(\ell_1,V_1,T_1)$ and $\mathbf p_2=(\ell_2,V_2,T_2)$.
\begin{itemize}
\item In the analysis of the speed heterogeneity (heterogeneity index $\delta_s$, cf.\ Figs.~\ref{fig2}, \ref{fig-cfmspeed}, \ref{fig-cfmspeed-difft0} and \ref{fig-cfmspeed-diffsig}) the time gap parameter $T$ ranges into $[0.05,1.95]$\,s by step of $0.05$\,s,
\begin{equation}
T_1=T+0.05\delta_s,\qquad T_2=T-0.05\delta_s,\qquad\qquad \delta_s=0,\ldots,19,
\end{equation}
while the desired (maximal) speed $V$ ranges into $[1,2]$\,m/s by step of $0.025$\,m/s,
\begin{equation}
V_1=V-0.025\delta_s,\qquad V_2=V+0.025\delta_s,\qquad\qquad \delta_s=0,\ldots,19.
\end{equation}
The parameters $\ell_1=\ell_2=\ell=0.3$\,m  for the agent size remain constant.
\item In the analysis of the size heterogeneity (heterogeneity index $\delta_l$, cf.\ Fig.~\ref{fig-cfmsize}) 
the parameter $\ell$ ranges into $[0,0.9]$\,m by step of $0.015$\,m decreasing and $0.03$\,m increasing,
\begin{equation}
\ell_1=\ell-0.015\delta_l,\qquad \ell_2=\ell+0.03\delta_l,\qquad\qquad \delta_l=0,\ldots,19.
\end{equation}
The remaining parameters for the agent speed $V_1=V_2=V=1.5$\,m/s and $T_1=T_2=T=1$\,s are constant.
\end{itemize}

\section*{Acknowledgements}
RK and AT acknowledge the Franco-German research project MADRAS funded in France by the Agence Nationale de la Recherche (ANR, French National Research Agency), project number ANR-20-CE92-0033, and in Germany by the Deutsche Forschungsgemeinschaft (DFG, German Research Foundation), project number 446168800. 

\section*{Data availability}
Simulations of lane or band formations with the two heterogeneity models $M_1$ Eq.~\ma\ and $M_2$ Eq.~\mb\ can be  implemented online and in real time on the dedicated web-page \href{https://www.vzu.uni-wuppertal.de/fileadmin/site/vzu/Lane-and-band-formation.html}{\texttt{\small https://www.vzu.uni-wuppertal.de/fileadmin/site/vzu/Lane-and- band-formation}}.
Simulation module and programming code (in Logo language) can be downloaded on the same link.

\section*{Competing interests}
The authors declare no competing financial interests.

\section*{Author contributions}
AS and AT conceived and designed the study. AT, BK and RK performed the numerical simulations. AT wrote the manuscript. All authors reviewed the article.

\newpage
\bigskip
\begin{center}
\LARGE\bf Supplementary Materials
\end{center}
\bigskip\bigskip

\noindent We use the collision-free (CF) model in the manuscript to analyse by simulation the two heterogeneity models $M_1$ Eq.~(1) and $M_2$ Eq.~(2). 
In these supplementary materials, we present identical simulation experiments as those carried out in the manuscript by using the Social Force (SF) model\cite{Helbing1995}. 
Generally speaking, similar phase transitions to lane and band patterns occur when using, respectively, the static heterogeneity models $M_1$ or the dynamic model $M_2$.
In the following, we first present the simulation results before providing technical details on the social force model.

\paragraph{Simulation results} 

\begin{figure}[!ht]\centering\bigskip\medskip
\centering 
\begin{minipage}[t]{.38\textwidth}
\begin{center}
\textbf{\small Model $M_1$} -- Static heterogeneity
\end{center}
\addpl{0}{0.24}{0.31}{0.33}{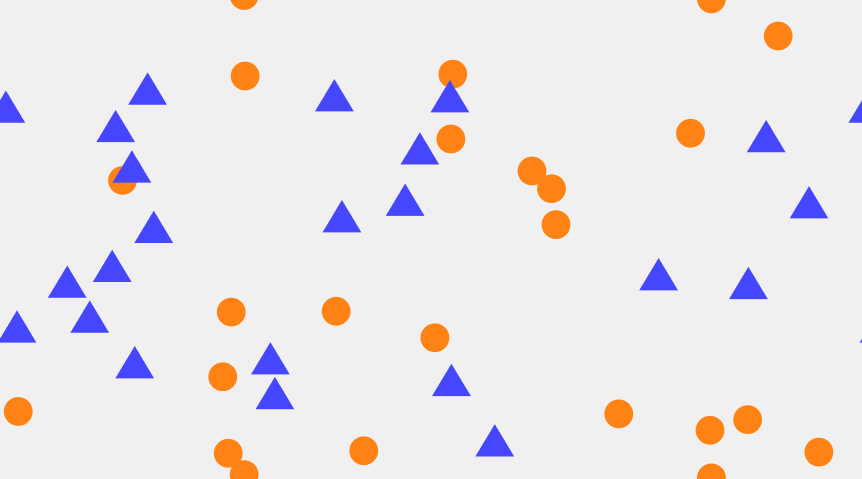}
\addpl{10}{1.39}{0.26}{0.19}{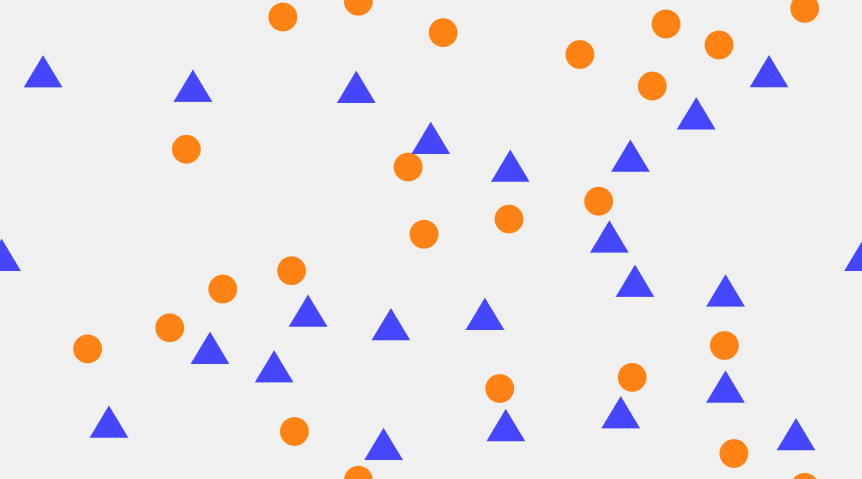}
\addpl{50}{1.27}{1}{0.12}{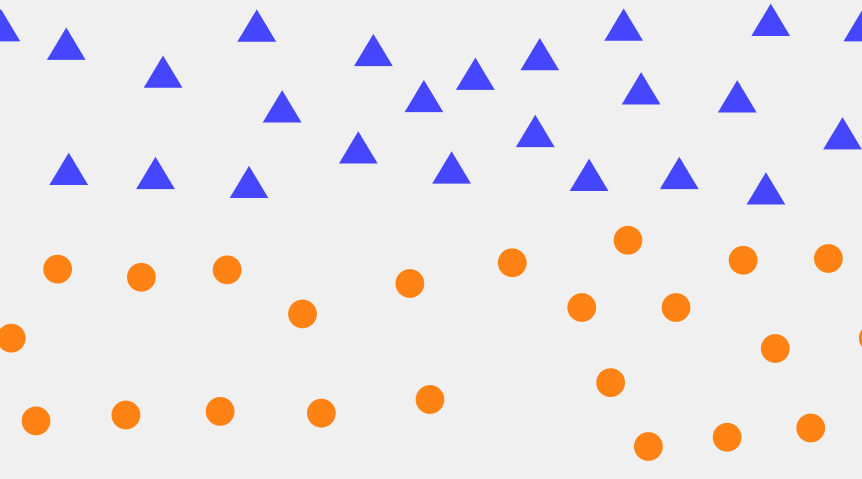}
\end{minipage}\hspace{2cm}\begin{minipage}[t]{.38\textwidth}
\begin{center}
\textbf{\small  Model $M_2$} -- Dynamic heterogeneity
\end{center}
\addpl{0}{0.2}{0.22}{0.1}{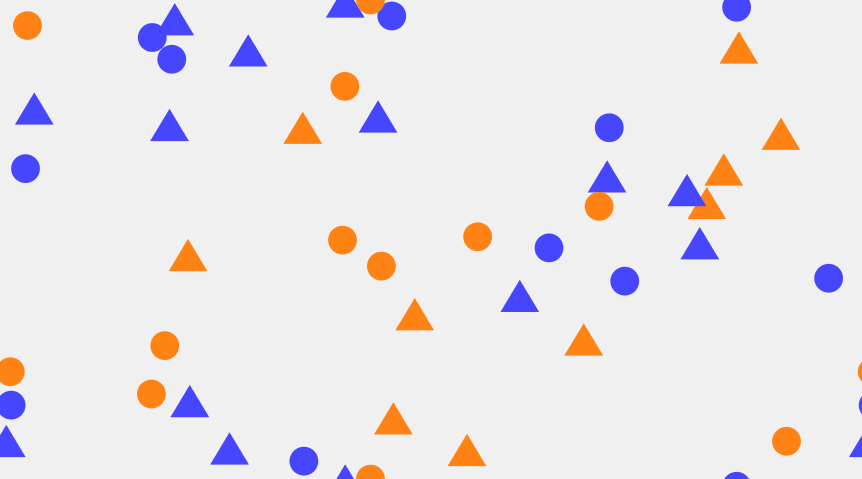}
\addpl{10}{1.27}{0.17}{0.21}{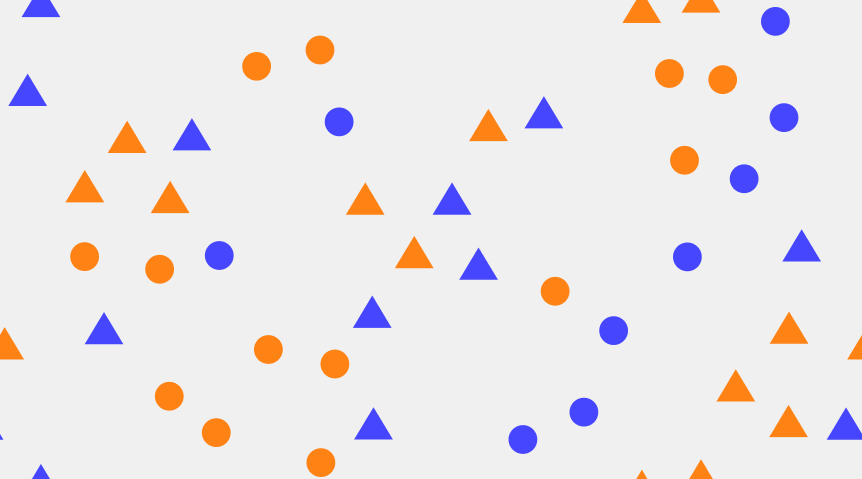}
\addpl{50}{1.34}{0.1}{1}{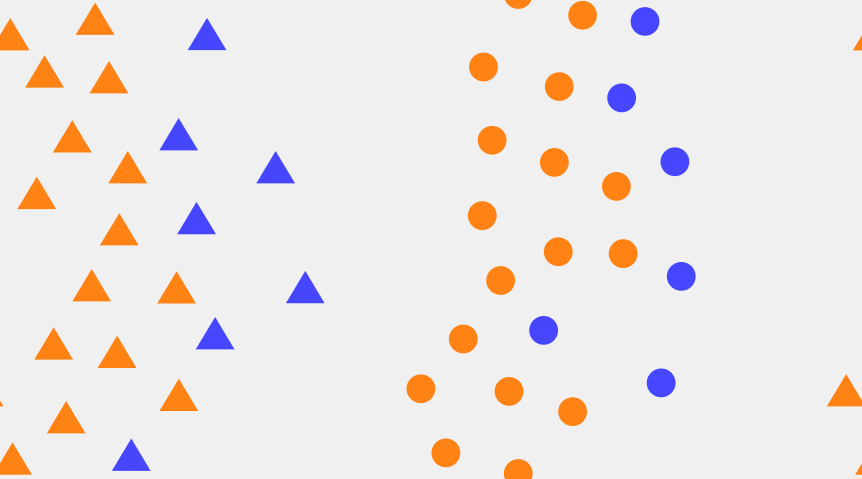}
\end{minipage}\\[-2mm]
\input{Figures/sfig1g.tex}
    \caption{Typical histories for the model $M_1$ with heterogeneity in the agent characteristics for which lanes emerge: the lane order parameter $\Phi_L$ tends to one while the band order parameter $\Phi_B$ is close to zero (left panels, $\delta_s=18$), and for the model $M_2$ with heterogeneity in the interactions where bands emerge, $\Phi_L$ is close to zero while $\Phi_B$ tends to one (right panels, $\delta_s=10$). SF motion model, flow direction from left to right, periodic boundary conditions, random initial conditions.}
    \label{sfig1}
\end{figure}

We simulate using the SF model the evolution of 45 agents on a $9\times5$~m rectangle with periodic boundaries (see Sec.~2 of the manuscript for details).
The preliminary experiment presents two system histories obtained with the two heterogeneity models for given heterogeneity indexes. 
As with CF model, we observe using the SF model rapid formation of lanes with the static heterogeneity type $M_1$, while band patterns emerge with the dynamic heterogeneity type $M_2$ (compare Fig.~2 in the manuscript to Fig.~\ref{sfig1}). 
The convergence to lane and band patterns is slower for larger systems (see the dotted curves obtained on a three times larger system with 135 agents on a $15\times9$~m rectangle).

\begin{figure}[!ht]
\centering\input{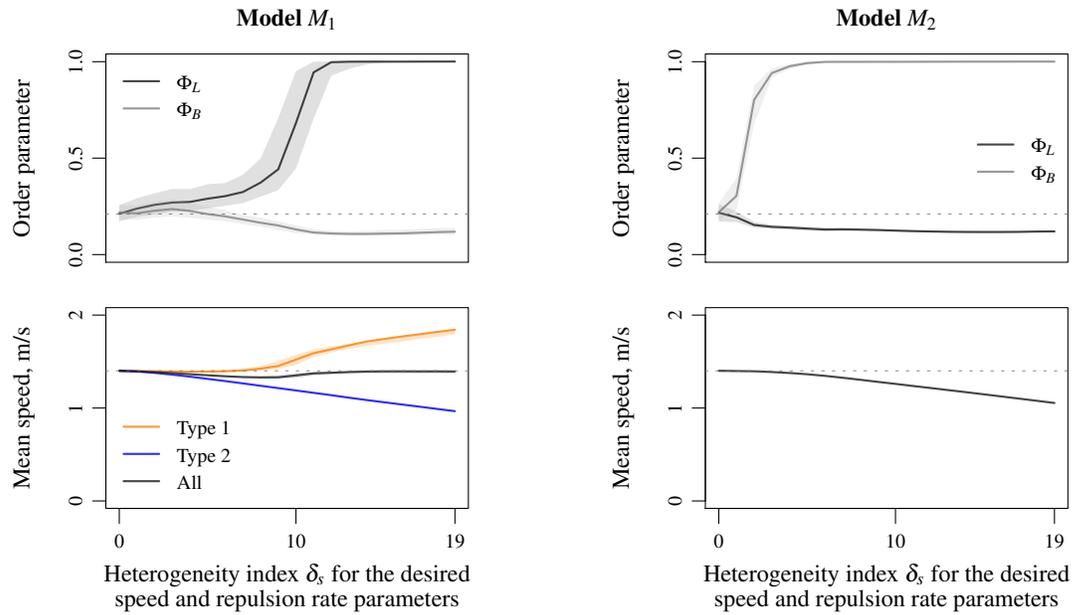}
\caption{Lane and band order parameters (top panels) and mean speed (bottom panels) according to the heterogeneity index of agent speed features for the static heterogeneity model $M_1$ (left panels) and for the dynamic heterogeneity model $M_2$ (right panels) with the SF model.
We observe qualitatively similar phase transitions as those obtained with the CF model, compare with Fig.~3 in the manuscript.}
\label{fig-sfmspeed}
\bigskip
\end{figure}

\begin{figure}[!ht]
\centering\input{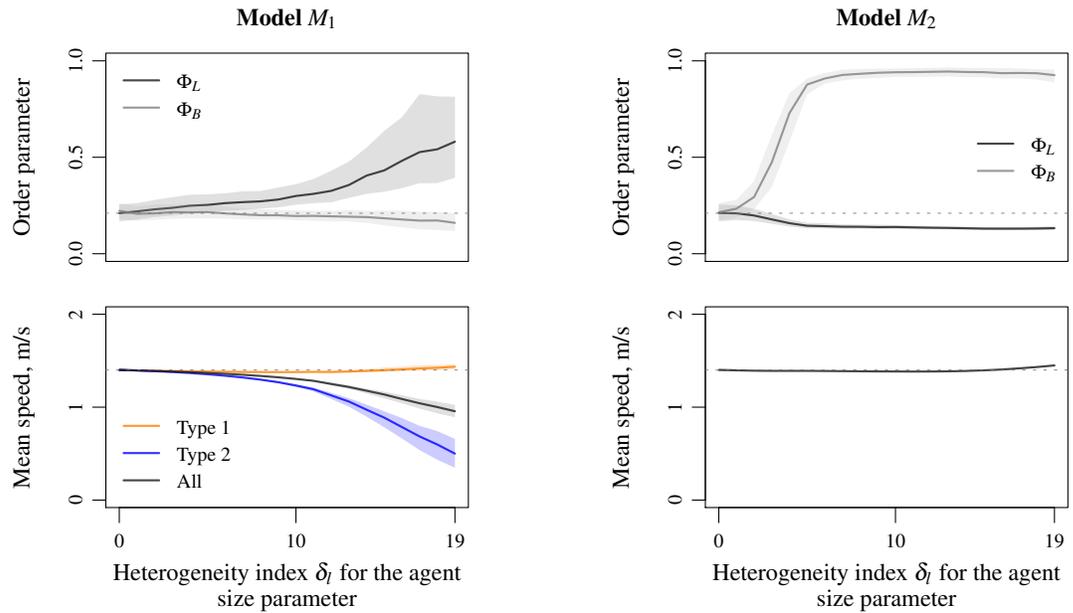}
\caption{Lane and band order parameters (top panels) and mean speed (bottom panels) according to the heterogeneity index of agent size for the SF model. In contrast to CF model (see Fig.~4 in the manuscript), the transition to lane patterns is slower in absence of strict exclusion rules between the agents.}
\label{fig-sfmsize}
\end{figure}

Phase transitions occur in stationary states as the heterogeneity indexes increase. 
The dynamics range from disorder states to ordered states with lane or band patterns and polarised order parameters, see Fig.~\ref{fig-sfmspeed}. 
This holds for heterogeneity relying on agent speed features (heterogeneity index $\delta_s$, see Fig.~\ref{fig-sfmspeed}) or on the agent size (heterogeneity index $\delta_l$, see Fig.~\ref{fig-sfmsize}). 
However, the transition to lane pattern is specially laborious when dealing with the size parameter with SF model (see Fig.~\ref{fig-sfmsize}, top left panel).  Indeed, the size parameter $\ell$ with SF model does not describe a hard-core exclusion between the agents as CF model does.
On the other hand. the phase transitions to band patterns (heterogeneity type $M_2$) occur for a lower heterogeneity feature again with SF model (compare Figs.~3 and 4 in the manuscript to Figs.~\ref{fig-sfmspeed} and \ref{fig-sfmsize}).

\begin{figure}[!ht]
\centering
\input{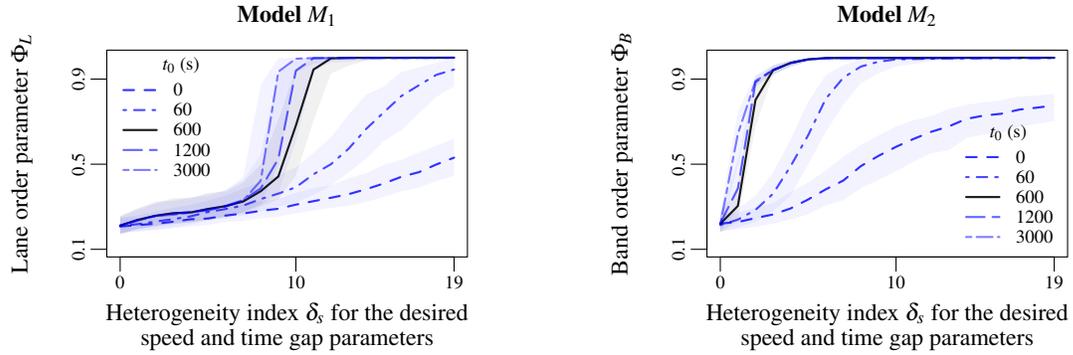}\smallskip
\caption{Lane order parameter for the static heterogeneity model $M_1$ (left panel), and band order parameter for the dynamic heterogeneity model $M_2$ (right panel) according to the heterogeneity index $\delta_s$ of agent speed features. 
The different curves correspond to different simulation times $t_0=0$, $60$, $600$, $1200$ and $3000$~s before starting the measurements (random initial conditions).
As for the CF model, the phase transition to lane and band patterns relatively fast emerges with the SF model. It can be observed during the first minutes of simulation. However, the system is not completely stabilised after 600~s.}
\label{fig-sfmspeed-difft0}
\end{figure}

\begin{figure}[!ht]
\centering
\input{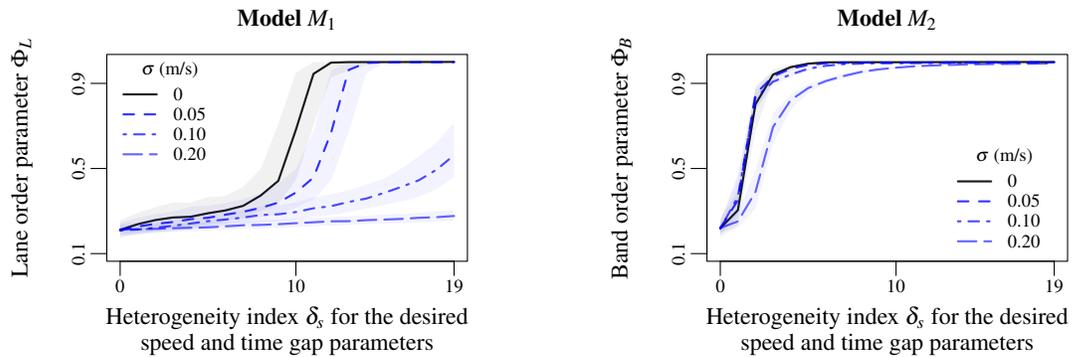}\smallskip
\caption{Lane order parameter for the static heterogeneity model $M_1$ (left panel), and band order parameter for the dynamic heterogeneity model $M_2$ (right panel) according to the heterogeneity index $\delta_s$ of agent speed features. 
The different curves correspond to different noise amplitudes $\sigma=0$, $0.1$, $0.2$ and $0.5$~m/s in the dynamics.
The noise clearly perturbs the lane formation (left panel, freezing-by-heating-effect).
Oppositely, the band formation is robust against the noise (right panel).}
\label{fig-sfmspeed-diffsig}
\end{figure}

The phase transitions to lane and band patterns can be observed during the first minutes of simulations (see Fig.~\ref{fig-sfmspeed-difft0}). 
However, in contrast to the results obtained with CF model, the system is not completely stabilised after 600~s, especially for the heterogeneity model $M_1$ (compare Fig.~5 in the manuscript to Fig.~\ref{fig-sfmspeed-difft0}). 
SF model, being an inertial model of the second order, requires longer simulation times to describe stationary performances.
Similarly to CF model, noising the dynamic clearly perturbs the lane formation (freezing-by-heating-effect, compare Fig.~6 in the manuscript to Fig.~\ref{fig-sfmspeed-diffsig}, left panels).
While, oppositely, the band formation is robust against the noise (see Fig.~\ref{fig-sfmspeed-diffsig}, right panel).

\paragraph{Definition of the social force model}

In the social force model, the dynamics of an agent $n$ with position $\mathbf x_n$ and neighborhood $\mathbf X_n$ is given by the second order differential equation
\begin{equation}
\ddot{\mathbf x}_n
= F^\textsc{\tiny sf}_{\mathbf p_j}(\mathbf X_n)
= \frac{1}{\tau}(V_j\,\mathbf e_0-\dot{\mathbf x}_n)+\sum_{m\not=n}\varphi(\mathbf e_{mn})\nabla U_j(\|\mathbf x_n-\mathbf x_m\|)+\sigma\xi_n.
\label{sfm}
\end{equation}
Here $V_j\ge0$ is the desired speed, $\ell_j\ge0$ the agent size, $\tau=0.5$~s a relaxation time, while $\varphi(\mathbf e)=1-\cos(\pi-\hat{\mathbf e})$ is the vision field factor with $\mathbf e_{mn}$ the direction from $m$ to $n$. As for CF model, $\mathbf e_0=0$ is the desired direction (polarity) and $U(x)=A_j\exp\big((\ell_j-x)/B\big)$ with parameters $A_j$ and $B=0.2$\,m is a repulsive potential with the neighbors.  A bi-dimensional white noise $\xi_n$ (i.e. the time derivatives of two independent Wiener processes) with amplitude $\sigma>0$ is used in Fig.~\ref{fig-sfmspeed-diffsig}.
The model is simulated using an explicit Euler numerical scheme in deterministic cases, and using an Euler-Maruyama scheme for the simulation including a stochastic noise. 
The time step is $\delta t=0.01$~s in both cases.

\paragraph{Setting of the parameters}

The default values for the parameters $\mathbf p=(\ell,V,A)$ of the SF model are based on the setting proposed in the literature\cite{Helbing1995}: $\ell=0.3$\,m, $V=1.5$\,m/s, and $A=3$\,m/s$^2$.
Starting from the default values, we vary using heterogeneity indexes the parameter settings $\mathbf p_1=(\ell_1,V_1,A_1)$ and $\mathbf p_2=(\ell_2,V_2,A_2)$.
\begin{itemize}
\item In the analysis of the speed heterogeneity (heterogeneity level $\delta_s$, cf.\ Figs.~\ref{sfig1}, \ref{fig-sfmspeed}, \ref{fig-sfmspeed-difft0} and \ref{fig-sfmspeed-diffsig}) the expulsion rate $A$ ranges into $[2,4]$\,m/s$^2$ by step of $0.05$\,m/s$^2$: $A_1=A-0.05\delta_s$, $A_2=A+0.05\delta_s$, $\delta_s=0,\ldots,19$.
The desired (maximal) speed $V$, as with CF model, ranges into $[1,2]$\,m/s by step of $0.025$\,m/s: $V_1=V-0.025\delta_s$, $V_2=V+0.025\delta_s$, $\delta_s=0,\ldots,19$.
Note that $\tau_j=V_j/A_j$ systematically holds for all $j=1,2$.
The parameter $\ell_1=\ell_2=\ell$ for the agent size remains constant.
\item In the analysis of the size heterogeneity (heterogeneity level $\delta_l$, cf.\ Fig.~\ref{fig-sfmsize}) 
the parameter $\ell$, as with CF model, ranges into $[0,0.9]$\,m by step of $0.015$~m decreasing and $0.03$\,m increasing: $\ell_1=\ell-0.015\delta_l$, $ \ell_2=\ell+0.03\delta_l$, $\delta_l=0,\ldots,19$.
The remaining parameters for the agent speed $V_1=V_2=V$, and $A_1=A_2=A$ are constant.
\end{itemize}
\end{document}